\documentstyle[aps,epsf]{revtex}

\begin{document}

\baselineskip 14pt

\newcommand{\sla}[1]{/\!\!\!#1}

\title{Searching for $VV \to H \to \tau\tau$ at the CERN LHC}
\author{D. Rainwater}
\address{University of Wisconsin - Madison}

\maketitle

\begin{abstract}
The production of a neutral, CP even Higgs via weak boson fusion and 
decay $H\to\tau\tau$ at the LHC is studied for the Standard Model and 
MSSM, utilizing a parton level Monte Carlo analysis. This channel allows 
a $5\sigma$ observation of a Standard Model Higgs with an integrated 
luminosity of about 30~fb$^{-1}$, and provides a direct measurement of 
the $H\tau\tau$ coupling. For the MSSM case, a highly significant signal 
for at least one of the Higgs bosons with reasonable luminosity is 
possible over the entire physical parameter space which will be left 
unexplored by LEP2.
\
\end{abstract}

\section{Introduction}

The search for the Higgs boson and, hence, for the origin of electroweak 
symmetry breaking and fermion mass generation, remains one of the premier 
tasks of present and future high energy physics experiments. Fits to 
precision electroweak (EW) data have for some time suggested a relatively 
small Higgs boson mass, of order 100~GeV~\cite{EWfits}, hence we have 
studied an intermediate-mass Higgs, with mass in the $110-150$~GeV range, 
beyond the reach of LEP at CERN and of the Fermilab Tevatron. Observation 
of the $H\to\tau\tau$ decay channel in weak boson fusion events at the 
CERN Large Hadron Collider (LHC) is quite promising, both in the Standard 
Model (SM) and Minimal Supersymmetric Standard Model (MSSM). This channel 
has lower QCD backgrounds compared to the dominant $H\to b\bar b$ mode, 
thus offering the best prospects for a direct measurement of a $Hf\bar f$ 
coupling.

Despite weak-boson fusion Higgs production's significantly lower 
cross section at the LHC (almost one order of magnitude), it has the 
advantage of additional information in the event other than the decay 
products' transverse momentum and their invariant mass resonance: the 
observable quark jets. Thus one can exploit techniques like 
forward jet tagging~\cite{Cahn,BCHP,DGOV} to reduce the backgrounds. 
Another advantage is the different color structure of the signal v. the 
background. Additional soft jet activity (minijets) in the 
central region, which occurs much more frequently for the color-exchange 
processes of the QCD backgrounds~\cite{bpz_minijet}, are suppressed via 
a central jet veto.

We have performed a first analysis of intermediate-mass SM $H\to\tau\tau$ 
and of the main physics and reducible backgrounds at the LHC, which 
demonstrates the feasibility of Higgs boson detection in this channel,
with modest luminosity~\cite{HRZ_tau}. $H\to\tau\tau$ event 
characteristics were analyzed for one $\tau$ decaying leptonically and 
the other decaying hadronically. We demonstrated that forward jet 
tagging, $\tau$ identification and reconstruction criteria alone yield 
an $\approx$2/1 signal-to-background (S/B) ratio; additional large 
background suppression factors can be obtained with the minijet veto.

In the MSSM, strategies to identify the structure of the Higgs sector are 
much less clear. For large tan$\beta$, the light neutral Higgs bosons may 
couple much more strongly to the $T_3 = -1/2$ members of the weak isospin 
doublets than its SM analogue. As a result, the total width can increase 
significantly compared to a SM Higgs of the same mass. This comes at the 
expense of the branching ratio $B(h\to\gamma\gamma)$, the cleanest Higgs 
discovery mode, possibly rendering it unobservable over much of MSSM 
parameter space and forcing consideration of other observational channels. 
Since $B(h\to\tau\tau)$ is instead slightly enhanced, we have examined the 
$\tau$ mode as an alternative~/cite{PRZ}.

\section{Simulations of Signal and Backgrounds}

The analysis uses full tree-level 
matrix elements for the weak boson fusion Higgs signal and the various 
backgrounds. Extra minijet activity was simulated by adding the emission 
of one extra parton to the basic signal and background processes, with 
the soft singularities regulated via a truncated shower approximation 
(TSA)~\cite{TSA}. 

We simulated $pp$ collisions at the CERN LHC, $\protect\sqrt{s} = 14$~TeV. 
For all QCD effects, the running of the strong-coupling constant was 
evaluated at one-loop order, with $\alpha_s(M_Z) = 0.118$. We employed 
CTEQ4L parton distribution functions~\cite{CTEQ4} throughout. The 
factorization scale was chosen as $\mu_f =$ min($p_T$) of the defined 
jets, and the renormalization scale $\mu_r$ was fixed by 
$(\alpha_s)^n = \prod_{i=1}^n \alpha_s(p_{T_i})$. 
Detector effects were considered by including Gaussian smearing for 
partons and leptons according to ATLAS expectations~\cite{CMS-ATLAS}.

At lowest order, the signal is described by two single-Feynman-diagram 
processes, $qq \to qq(WW,ZZ) \to qqH$, i.e. $WW$ and $ZZ$ fusion where the 
weak bosons are emitted from the incoming quarks~\cite{qqHorig}. From a 
previous study of $H\to\gamma\gamma$ decays in weak boson 
fusion~\cite{RZ_gamma} we know several features of the signal, which we 
could directly exploit here: the centrally produced Higgs boson tends to 
yield central decay products (in this case $\tau^+\tau^-$), and the two 
quarks enter the detector at large rapidity compared to the $\tau$'s 
and with transverse momenta in the 20-80 GeV range, thus leading to two 
observable forward tagging jets.

We only considered the case of one $\tau$ decaying leptonically 
($e$,$\mu$), and the other decaying hadronically, with a combined 
branching fraction of $45\%$. Our analysis critically employed transverse 
momentum cuts on the charged $\tau$-decay products and, hence, some care 
was taken to ensure realistic momentum distributions. Because of its small 
mass, we simulated $\tau$ decays in the collinear and narrow-width 
approximations and with decay distributions to $\pi$,$\rho$,
$a_1$~\cite{HMZ}, adding the various hadronic decay modes according to 
their branching ratios. We took into account the anti-correlation of the 
$\tau^\pm$ polarizations in the decay of the Higgs.

Positive identification of the hadronic $\tau^\pm\to h^\pm X$ decay requires
severe cuts on the charged hadron isolation. We based our simulations on the 
possible strategies analyzed by Cavalli {\it et al.}~\cite{Cavalli}. 
Considering hadronic jets of $E_T>40$~GeV in the ATLAS detector, they found 
non-tau rejection factors of 400 or more while true hadronic $\tau$ decays 
are retained with an identification efficiency of $26\%$.

Given the H decay signature, the main physics background to the 
$\tau^+ \tau^- jj$ events of the signal arises from real emission QCD 
corrections to
the Drell-Yan process $q\bar{q} \to (Z,\gamma) \to \tau^+\tau^-$, dominated 
by $t$-channel gluon exchange. All interference effects between virtual 
photon and $Z$-exchange were included, as was the correlation of $\tau^\pm$ 
polarizations. The $Z$ component dominates, so we call these processes 
collectively the ``QCD $Zjj$'' background.

An additional physics ``EW $Zjj$'' background arises from $Z$ and $\gamma$ 
bremsstrahlung in (anti)quark scattering via $t$-channel electroweak boson 
exchange, with subsequent decay $Z,\gamma\to \tau^+\tau^-$. Naively, this 
EW background may be thought of as suppressed compared to the analogous QCD 
process. However, the EW background includes electroweak boson fusion, 
$VV \to \tau^+\tau^-$, which has a momentum and color structure identical 
to the signal and thus cannot easily be suppressed via cuts.

Finally, we considered reducible backgrounds, i.e. any event that can mimic 
the $Hjj$ signature of a hard, isolated lepton and missing $p_T$, a hard, 
narrow $\tau$-like jet, and two forward tagging jets. Thus we examined 
$W+jets$, where the $W$ decays leptonically ($e$,$\mu$) and one jet fakes 
a hadronic $\tau$, and $b\bar b+jets$, where one $b$ decays leptonically 
and either a light quark or $b$ jet fakes a hadronic $\tau$. We neglected 
other sources like $t\bar t$ events which had previously been shown to give 
substantially smaller backgrounds~\cite{Cavalli}. 

Fluctuations of a parton into a narrow $\tau$-like jet are considered with 
probability $0.25\%$ for gluons and light-quark jets and $0.15\%$ for $b$ 
jets (which may be considered an upper bound)~\cite{Cavalli}.

In the case of $b\bar b+jj$, we simulated the semileptonic decay 
$b\to\ell\nu c$ by multiplying the $b\bar{b}jj$ cross section by a 
branching factor of 0.395 and implementing a three-body phase space 
distribution for the decay momenta to estimate the effects of lepton 
isolation cuts. We normalized our resulting cross section to reproduce the 
same factor 100 reduction found in Ref.~\cite{Cavalli}.

\section{Standard Model Analysis}

The basic acceptance requirements must ensure that the two jets and two 
$\tau$'s are observed inside the detector (within the hadronic and 
electromagnetic calorimeters, respectively), and are well-separated from each 
other:
\begin{eqnarray}
\label{eq:cuts}
& p_{T_{j(1,2)}} \geq 40, 20~{\rm GeV} \, ,\qquad |\eta_j| \leq 5.0 \, ,\qquad 
\triangle R_{jj} \geq 0.7 \, , & \nonumber\\
& |\eta_{\tau}| \leq 2.5 \, , \qquad \triangle R_{j\tau} \geq 0.7 \, , \qquad
\triangle R_{\tau\tau} \geq 0.7 \, . &
\end{eqnarray}

The $Hjj$ signal is characterized by two forward jets with large invariant 
mass, and central $\tau$ decay products. The QCD backgrounds have a large 
gluon-initiated component and thus prefer lower invariant tagging jet masses. 
Also, their $\tau$ and $W$ decay products tend to be less central. Thus, to 
reduce the backgrounds to the level of the signal we required tagging jets 
with a combination of large invariant mass, far forward rapidity, and high 
$p_T$, and $\tau$ decay products central with respect to the tagging 
jets~\cite{RZ_gamma}:
\begin{eqnarray}
\eta_{j,min} + 0.7 < \eta_{\tau_{1,2}} < \eta_{j,max} - 0.7 \, , \qquad
\eta_{j_1} \cdot \eta_{j_2} < 0 \, \\
\triangle \eta_{tags} = |\eta_{j_1}-\eta_{j_2}| \geq 4.4 \, , \qquad
m_{jj} > 1 \, {\rm TeV} \, .
\end{eqnarray}

Triggering the event via the isolated $\tau$-decay lepton and identifying the 
hadronic $\tau$ decay as discussed in Ref.~\cite{Cavalli} requires sizable 
transverse momenta for the observable $\tau$ decay products:
$p_{T_{\tau,lep}} > 20~{\rm GeV}$ and $p_{T_{\tau,had}} > 40~{\rm GeV}$. It 
is possible to reconstruct the $\tau$-pair invariant mass from the 
observable $\tau$ decay products and the missing transverse momentum vector 
of the event~\cite{tautaumass}. The $\tau$ mass was neglected and collinear 
decays assumed, a condition easily satisfied because of the high $\tau$ 
transverse momenta required. The $\tau$ momenta were reconstructed from the 
charged decay products' $p_T$ and missing $p_T$ vectors. We imposed a cut on 
the angle between the $\tau$ decay products to satisfy the collinear decay 
assumption, $cos(\theta_{\ell h}) > -0.9$, and demanded a physicality 
condition for the reconstructed $\tau$ momenta (unphysical solutions arise 
from smearing effects); that is, the fractional momentum $x_{\tau}$ a charged 
decay observables takes from its parent $\tau$ cannot be negative. 
Additionally, the $x_{\tau_{\ell}}$ distribution of the leptonically decaying 
$\tau$-candidate is softer for real $\tau$'s than for the reducible 
backgrounds, because the charged lepton shares the parent $\tau$ energy with 
two neutrinos. A cut $x_{\tau_l} < 0.75$, $x_{\tau_h} < 1$ proved very 
effective in suppressing the reducible backgrounds.

Our Monte Carlo predicted a $\tau$-pair mass resolution of 10 GeV or better, 
so we choose $\pm 10$~GeV mass bins for analyzing the cross sections. 
Finally, the $Wj+jj$ background exhibits a Jacobian peak in its $m_T$ 
distribution~\cite{Cavalli}. A cut $m_T(\ell,\sla{p_T}) < 30$~GeV largely 
eliminates this background.

Using all these cuts together, although not in a highly optimized 
combination, we expect already a signal to background ratio of 2/1 with a 
signal cross section of 0.5 fb for $m_H = 120$~GeV.

A probability for vetoing additional central hadronic radiation was 
obtained by measuring the fraction of events that have additional radiation 
in the central region, between the tagging jets, with $p_T$ above 20 GeV, 
using the matrix elements for additional parton emission. This minijet veto 
reduces the signal by about $30\%$, but eliminates typically $85\%$ of the 
QCD backgrounds; the EW $Zjj$ background is reduced by about $50\%$, 
indicating the presence of both boson bremsstrahlung and weak boson fusion 
effects. Because the veto probability for QCD backgrounds is found to be 
process independent, we applied the same value to the $bb+jj$ background. 
Table~\ref{table1} gives expected numbers of events for 30 fb$^{-1}$ 
integrated luminosity at the LHC.

\begin{figure}[htb]
\begin{picture}(0,0)(0,0)
\includegraphics{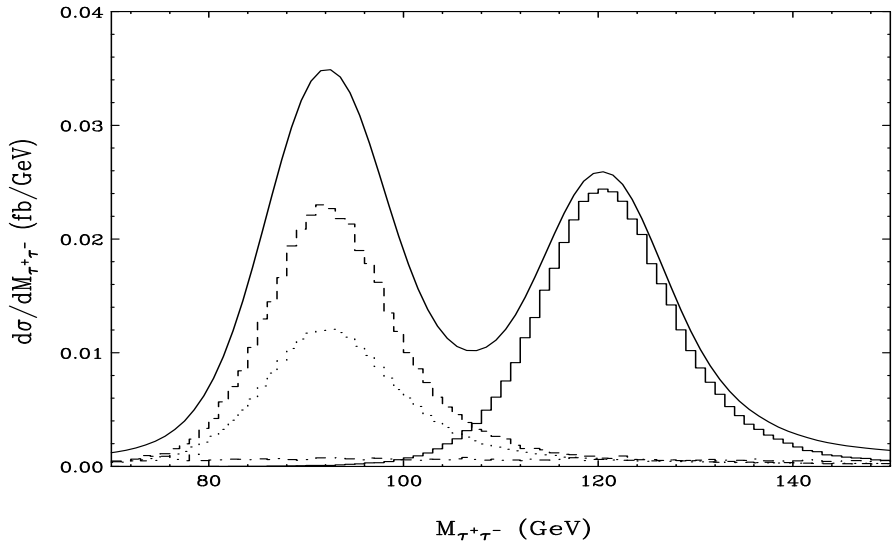}
\end{picture}
\vspace{7.2 cm}
\caption[]{
\label{fig:Mtautau}
\small Reconstructed $\tau$ pair invariant mass distribution for the 
signal and backgrounds after all cuts and multiplication by the 
expected survival probabilities. 
The solid line represents the sum of the signal and all backgrounds.
Individual components are shown as histograms:
the $Hjj$ signal (solid), the irreducible QCD $Zjj$
background (dashed), the irreducible EW $Zjj$ background (dotted), and the
combined $Wj+jj$ and $b\bar{b}jj$ reducible backgrounds (dash-dotted).}
\end{figure}

It is possible to isolate a virtually background-free 
$qq \to qqH \to jj\tau\tau$ signal at the LHC, leading to a $5\sigma$ 
observation of a SM Higgs boson with a mere 30 fb$^{-1}$ of data. The 
expected purity of the signal is demonstrated in Fig.~\ref{fig:Mtautau}, 
showing the reconstructed $\tau\tau$ invariant mass for a SM Higgs of 120 GeV 
after all cuts and a minijet veto have been applied. While the reducible 
$Wj+jj$ and $b\bar b+jj$ backgrounds are the most complicated and do require 
further study, they appear to be easily manageable. 

\begin{table}
\caption[]{\label{table1}
Number of expected events for the signal and backgrounds, for 
30~${\rm fb}^{-1}$ integrated luminosity, all cuts and application of a 
minijet veto. Mass bins are $\pm$10 GeV. Gaussian equivalent significances 
are stated, based on a proper Poisson statistical analysis. 
From~\cite{HRZ_tau}.}
\begin{tabular}{c|ccccc|c}
$m_H$(GeV) & $Hjj$ & QCD $Zjj$
       & EW $Zjj$ & $Wj+jj$ & $\;\;b\bar{b}jj\;\;$ & $\sigma_{Gauss}$ \\
\hline
110 & 11.1 & 2.1  & 1.4  & 0.1 & 0.3 & 4.1 \\
120 & 10.4 & 0.6  & 0.5  & 0.1 & 0.2 & 5.2 \\
130 &  8.6 & 0.3  & 0.3  & 0.1 & 0.2 & 5.0 \\
140 &  5.8 & 0.2  & 0.2  & 0.1 & 0.2 & 3.9 \\
150 &  3.0 & 0.1  & 0.2  & 0.1 & 0.2 & 2.3 \\
\end{tabular}
\end{table}

\section{MSSM Analysis}

The production of CP even Higgs bosons in weak boson fusion is governed by 
the $hWW,HWW$ couplings, which, compared to the SM case, are suppressed by 
factors $sin(\beta-\alpha),cos(\beta-\alpha)$, respectively~\cite{mssm}. 
Their branching ratios are modified with slightly more complicated factors. 
One can simply multiply SM cross section results from our analysis by 
these factors to determine the observability of $H \to \tau\tau$ in MSSM 
parameter space. We used a renormalization group improved next-to-leading 
order calculation, which allows a light Higgs mass up to $\sim 125$~GeV, 
and examined two trilinear term mixing cases, no mixing and maximal 
mixing~\cite{PRZ}.

Varying the pseudoscalar Higgs boson mass $m_A$, one finds that $m_h$,$m_H$ 
each approach a plateau for the case $m_A \to \infty,0$, respectively. 
Below $m_A \sim 120$~GeV, the light Higgs mass will fall off linearly with 
$m_A$, while the heavy Higgs will approach $m_H \sim 125$~GeV, whereas 
above $m_A \sim 120$~GeV, the light Higgs will approach $m_h \sim 125$~GeV 
and the heavy Higgs mass will rise linearly with $m_A$. The transition 
region behavior is very abrupt for large $tan\beta$, such that the plateau 
state will go to $\sim 125$~GeV almost immediately, while for small 
$tan\beta$ the transition is much softer and the plateau state reaches the 
limiting value via a more gradual asymptotic approach. 

\begin{figure}[htb] 
\centerline{\epsfxsize 2.8 truein \epsfbox{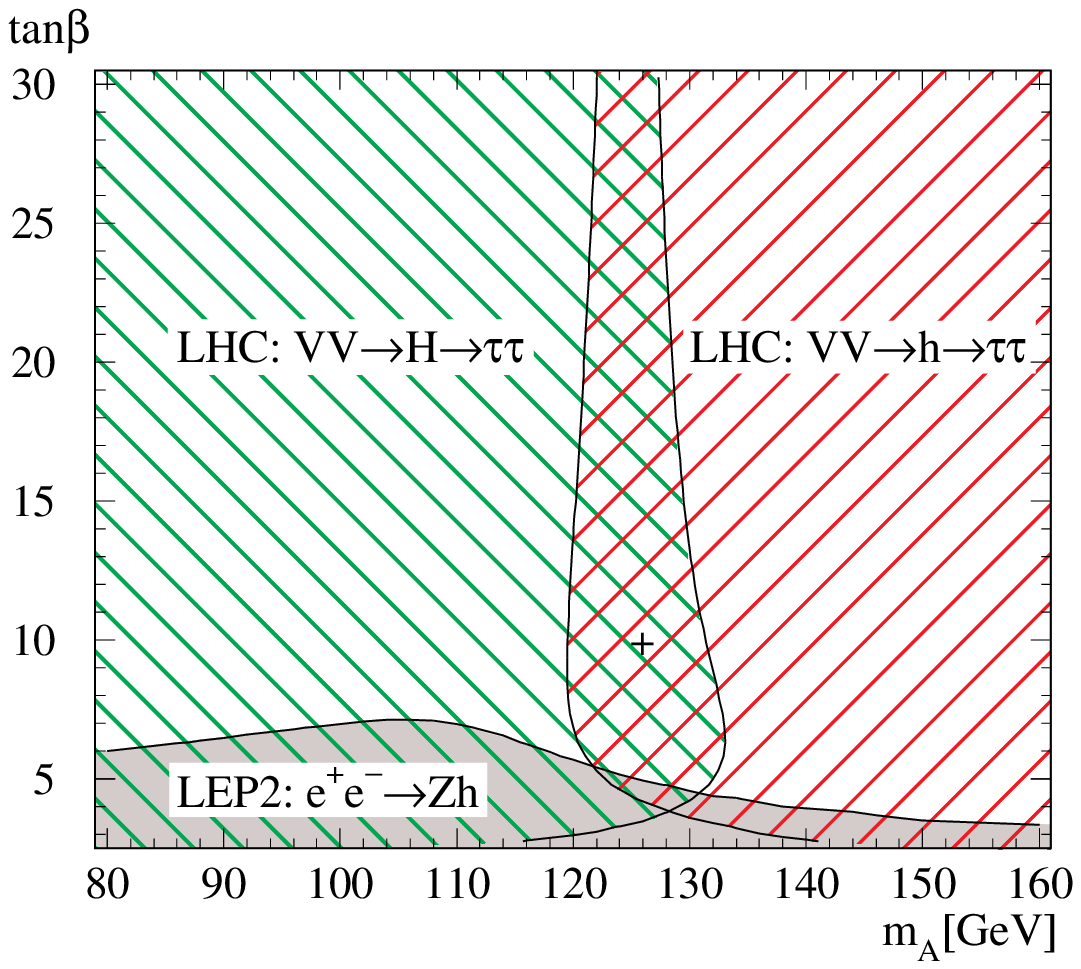}  
  \hspace{0.5in} \epsfxsize 2.8 truein \epsfbox{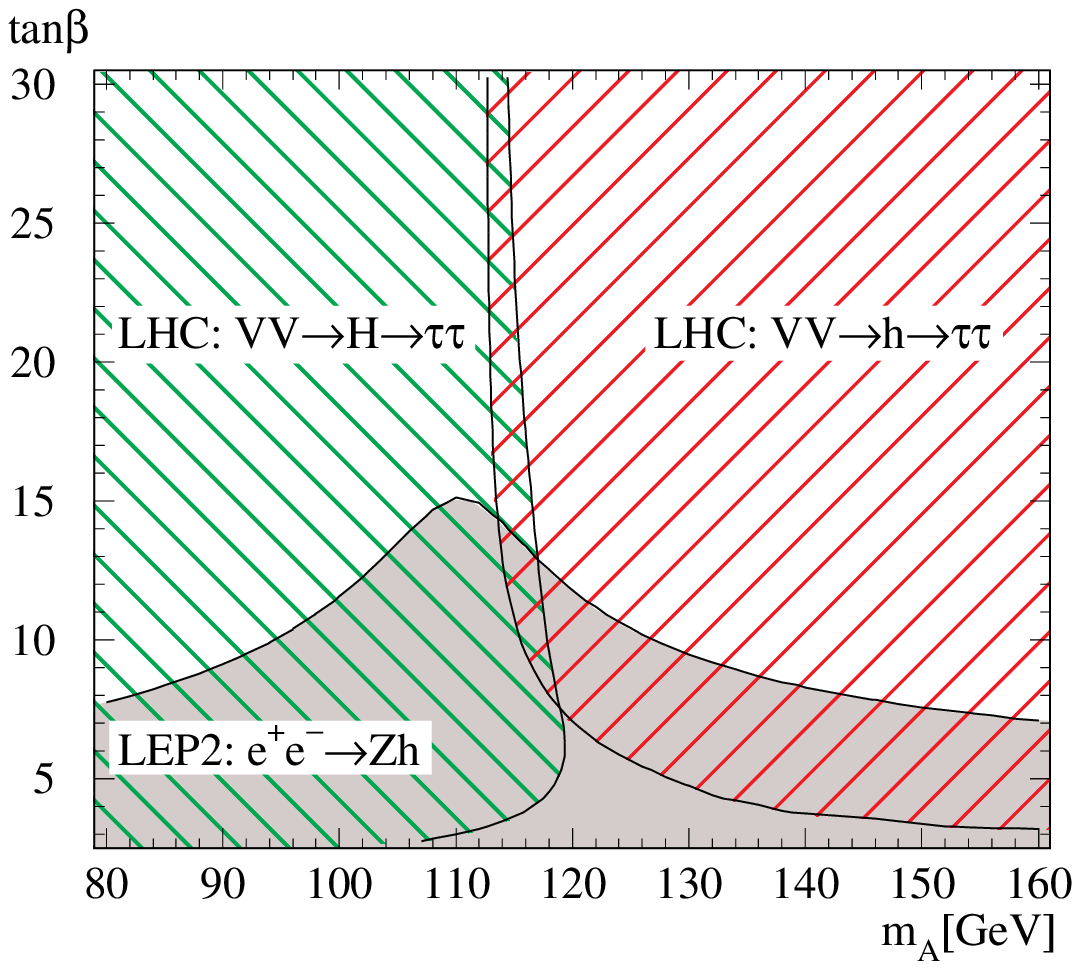} }
\vspace{0.2in}
\caption[]{\label{MSSM}
$5\sigma$ discovery contours for $h\to\tau\tau$ and $H\to\tau\tau$ in
weak boson fusion at the LHC, with $100$~fb$^{-1}$. Also shown are the
projected LEP2 exclusion limits. Results are shown for maximal mixing 
(left) and no mixing (right). From~\cite{PRZ}.}
\end{figure}

With reasonable luminosity, 100 fb$^{-1}$ in the worst case, it will be 
possible to observe at the $5\sigma$ level either $h$ or $H$ decays to 
$\tau$ pairs when they are in their respective plateau region, with the 
possibility of some overlap in a small region of $m_A$, as shown in 
Fig.~\ref{MSSM}. Very low values of $tan(\beta)$ would be unobservable, but 
already excluded by LEP2; there should be considerable overlap between this 
mode at the LHC and the LEP2 excluded region. Furthermore, a parton shower 
Monte Carlo with full detector simulation should be able to optimize the 
analysis so that much less data is required to observe or exclude the MSSM 
Higgs.

\section{Conclusions}

We have shown that weak boson fusion production of a Higgs boson with 
subsequent decay to $\tau$ pairs, $qq \to qqH \to jj\tau\tau$, is an 
excellent observational mode at the LHC, requiring only modest integrated 
luminosity and allowing direct measurement of the $H\tau\tau$ coupling. 
This mode also provides a no-lose strategy for seeing at least one of the 
CP even neutral MSSM Higgs bosons.

%
%

\end{document}